\documentclass{llncs}
\usepackage[utf8]{inputenc}
\usepackage[colorinlistoftodos]{todonotes}
\usepackage{parskip,amssymb,multicol,amsmath,float,caption,url}
\usepackage[ruled,noline]{algorithm2e}
\usepackage{comment}
\newcommand{\suchthat}{\mathbin{\mid}}

\usepackage{tikz}
\usetikzlibrary{positioning}

\title{Link Prediction using Top-$k$ Shortest Distances}

\begin{document}
\author{Andrei Lebedev\and
        JooYoung Lee\and
        Victor Rivera\and 
        Manuel Mazzara 
        }
 \institute{Institute of Information Systems, Innnopolis University, Russia
 \\
 \email{\{a.lebedev, j.lee\}@innopolis.ru},
 \and
 Institute of Technologies and Software Development, Innopolis University, Russia 
 \\
 \email{\{v.rivera, m.mazzara\}@innopolis.ru}}

\institute{Innnopolis University, Russia
\\
\email{\{a.lebedev, j.lee, v.rivera, m.mazzara\}@innopolis.ru},
}


\maketitle
\thispagestyle{empty}
\begin{abstract}
In this paper, we apply an efficient top-$k$ shortest distance routing algorithm to the link prediction problem and test its efficacy. We compare the results with other base line and state-of-the-art methods as well as with the shortest path. Our results show that using top-$k$ distances as a similarity measure outperforms classical similarity measures such as Jaccard and Adamic/Adar.   




\end{abstract}


\section{Introduction}
In a connected world, emphasis is on relationships more than isolated pieces of information. Relational databases may compute relationships at query time \cite{Qu:2014,Qu:2015}, but this would result in computationally expensive solution. 
Graph databases \cite{GDB:2008} store connections as first class citizens, allowing access to persistent connections in almost constant-time \cite{Robinson:2013}. One of the fundamental topological feature in the context of graph theory and graph databases, with implications in Artificial Intelligence and Web communities \cite{Goldberg:2005:CSP:1070432.1070455}, is the computation of the shortest-path distance between vertices \cite{lee2014estimating}.


Many efficient methods for searching shortest path were proposed. On the other hand, top-$k$ distance query handling methods are not well developed and spread. They have many advantages over traditional shortest path as they reveal much more information over a simple shortest path \cite{AAAI159320}. To extract top-$k$ distances from graph databases, an efficient indexing algorithm is needed, such as pruned landmark labeling scheme, presented in \cite{Akiba:2013:FES:2463676.2465315}. We utilize this algorithm to obtain the distances and then develop a similarity metric based on them to predict links on graphs.

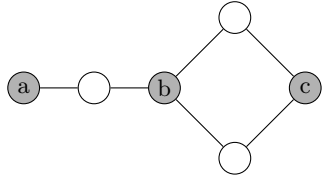
\begin{figure}[h]
	\centering
	\begin{tikzpicture}[node distance=5mm,
        bnode/.style={circle,draw=black,fill=black!30,minimum size=12pt,inner sep=0pt},
        gnode/.style={circle,draw=black,minimum size=12pt,inner sep=0pt},
		enode/.style={circle,minimum size=12pt,inner sep=0pt}
        ]
    \node[bnode](b){b};
    
    \node[enode](g)[right=of b]{};
    
    \node[gnode](e)[above=of g]{};
    \node[gnode](f)[below=of g]{};
    \node[bnode](c)[right=of g]{c};
    
    \node[gnode](d)[left=of b]{};
    \node[bnode](a)[left=of d]{a};
    
    \draw[-](a)to(d);
    \draw[-](d)to(b);
    \draw[-](b)to(e);
    \draw[-](b)to(f);
    \draw[-](e)to(c);
    \draw[-](f)to(c);
    
	\end{tikzpicture}
                     \caption{An example of representing connection between pairs of vertices using top-$k$ shortest distance.}   
 	\label{ex}   
\end{figure}

Figure \ref{ex} is an example to demonstrate the superiority of using top-$k$ distance over a simple distance metric to measure relationship between two vertices.
Table \ref{table} summarizes the connections represented in Figure \ref{ex}. Based on (top-$1$) distance alone, we can conclude that similarity (the shortest distance) between black nodes ($\{a,b\},\{b,c\}$) is the same in the graph. However, it is clear that $\{b,c\}$ are connected more tightly through a greater number of shortest paths.

The framework described in this paper is based on previous work which is called Pruned Landmark Labeling \cite{Akiba:2013:FES:2463676.2465315}. We evaluate and test the proposed algorithm extensively to prove the efficacy of the algorithm. Then we apply the algorithm to the link prediction problem to compare with existing solutions.


\begin{table}[H]
\centering
\caption{Distances and top-$k$ distances between pairs of black vertices in Figure \ref{ex}.}
\label{table}
\begin{tabular}{c|c|c}
\hline
\textbf{Vertex Pair} & \textbf{(Top-1) Distance} & \textbf{Top-$k$ Distance}\\ \hline
(a,b)   & 2                & {[}2, 4, 6\ldots {]} \\
(b,c)   & 2                & {[}2, 2, 4 \ldots{]}  \\ \hline
\end{tabular}
\end{table}

\section{Related Work}
In this section, we introduce some of comparable methods for computing top-$k$ shortest paths and link prediction as we utilize top-$k$ distances to predict links in social graphs.

\subsection{Top-k shortest paths}
One of the attempts to find $k$ shortest paths is presented in  \cite{Eppstein:1998:FSP} which achieves $\mathcal{O}(m+n\log{} n + k)$ complexity. \cite{Eppstein:1998:FSP} also covers programming problems, including the knapsack problem, sequence alignment, maximum inscribed polygons, and genealogical relationship discovery. We adopt the algorithm presented in \cite{AAAI159320} to discover $k$ shortest paths since it achieves six orders of magnitude faster computation given very large graphs with millions of vertices and tens of millions of edges. 

\subsection{Link prediction}
Link prediction in social network is a well known problem and extensively studied. Links are predicted using either semantic or topological information of a given network. The main idea in link prediction problem is to measure similarity between two vertices which are not yet linked to each other. If the measured similarity is high enough then the future link is predicted. \cite{ASI:ASI20591} attempts  to infer which new interactions among members of a social network are likely to occur in the near future. Authors develop approaches to link prediction based on measures for analyzing the ``proximity'' of nodes in a network. 

\section{Prediction Methods} 


Given a graph $G = (V,E)$, where $V$ is the set of vertices and $E$ is the set of edges. Then let $m$ and $n$ be $|E|$ and $|V|$ respectively. We also assume that vertices are represented by unique integers to enable the comparison of two vertices. Furthermore, let us denote $\mathcal{P}_{\mathrm{ab}}$ as the set of paths from  $a$ to $b$, $a \in V$ and $b \in V$, and $d_{i^{th}}(s,t)$ as the $i$-th shortest path between $s$ and $t$ in $\mathcal{P}_{\mathrm{st}}$.

In the following, we introduce the structure of querying and indexing algorithm from \cite{AAAI159320}.  

\begin{itemize}
\item \textit{Distance label $L(v)$}: a set of triplets $(u,\delta, c)$ of a vertex, a path length and a number of paths with length $\delta$.

\item \textit{Loop label $C(v)$}: a sequence of $k$ integers $(\delta_1,\delta_2,...,\delta_k)$.

\item \textit{Index $I = (L,C)$}: $L$ and $C$ sets of distance labels and loop labels.

\item \textit{Ordering}: vertices in order of decreasing degrees.

\item \textit{Number of paths} $c_{w,\delta'}$: number of paths in the $L(v)$ between vertex $v$ and $w$, of length not exceeding $\delta'$ using loop label $C(v)$.

\item Query$(I,s,t)$: smallest k elements in the $\Delta(I,s,t)$.
\end{itemize}
Then we can compute the multiset as follows. 
\begin{multline*}
\Delta(I,s,t) = \{\delta_{sv}+\delta_{vv}+\delta_{vt} \suchthat (v, \delta_{sv}) \in L(s),  \delta_{vv}\in C(v), (v, \delta_{vt}) \in L(t)\}
\end{multline*}

Referring to the original work by \cite{AAAI159320}, we have measured the performance in two ways: index construction speed and the final index size. Our implementation which considers unweighted and undirected graphs have achieved a reduction in index size compared with \cite{AAAI159320}.

{\bf Proposed Method:} First, we implemented the algorithm presented in \cite{AAAI159320} to compute top-$k$ shortest paths between two vertices. Then we use the sum of top-$k$ shortest paths as the similarity measure to predict future links. This naive approach shows better results compared with other commonly used link prediction methods. 
\begin{equation}
\label{eq}
S_k = \Sigma_{i=0}^{k-1} KSP(s,t,k)[i]
\end{equation}
Equation \ref{eq} shows the similarity measure based on top-$k$ distances where $KSP(s,t,k)$ is the list of top-$k$ shortest paths between vertices $s$ and $t$.

\section{Experimental Results}
{\bf Setup:}
In these experiments, all networks were treated as undirected unweighted graphs without self-loops and multiple edges. For testing purposes,  we randomly sample $60\%$ edges for prediction and $40\%$ for evaluation. The sampling, prediction and evaluation tasks were performed 10 times on each dataset. We use AUROC (Area Under the ROC curve) as an evaluation metric. We used five different datasets from \cite{AAAI159320} to ease comparisons.

{\bf Results:}
The performance of our proposed method is summarized in Table \ref{results}. The best performance for each dataset is emphasized. As we can see, Top-$4$ shortest distance consistently performs better than others except {\it CondMat} in which the difference is negligible. Intuitively, one might think bigger $k$ should be a better predictor but our results suggests that small $k$ is enough. Even with such a small $k$ value, 4, we can predict future links more effectively. From the experiments, we can demonstrate that top-$k$ distances capture the structural similarity between vertices better than commonly used measures, namely, {\it Common Neighbors (CN), Jaccard, Adamic/Adar} and {\it Preferential Attachment.}
\begin{table}[h!]
\centering
\caption{Performance evaluation of predictions on 5 datasets. Statistics of each graph are described in \cite{AAAI159320}.}
\label{results}
\hspace*{-.5cm}
\begin{tabular}{|l|c|c|c|c|c|c|c|c|}
\hline
           & Top1              & Top4              & Top16    & Top64    & CN       & Jaccard  & Adamic   & Preferential \\ \hline
Facebook-1 & 0.878481          & \textbf{0.909458} & 0.899959 & 0.886813 & 0.834086 & 0.833845 & 0.799192 & 0.693485     \\
Last.fm    & 0.863925          & \textbf{0.88316}  & 0.881586 & 0.87527  & 0.736351 & 0.733341 & 0.721282 & 0.782929     \\
GrQc       & 0.853479          & \textbf{0.853527} & 0.851746 & 0.84664  & 0.784875 & 0.784865 & 0.726173 & 0.720467     \\
HepTh      & 0.826561          & \textbf{0.82677}  & 0.824997 & 0.819686 & 0.730995 & 0.730984 & 0.671017 & 0.690331     \\
CondMat    & \textbf{0.911328} & 0.911099          & 0.90812  & 0.901543 & 0.815252 & 0.815241 & 0.75657  & 0.716568    \\
\hline
\end{tabular}
\end{table}
\section{Conclusions and Future work}
In this paper, we defined a new similarity metric between two users of social networks based on top-$k$ shortest distances. We also found out through experiments that our simple metric outperforms other common metrics and also a small $k$ suffices to accurately predict future links. Since top-$k$ distances capture important topological properties between vertices, we plan to apply the metric in gene regulation networks to discover unknown  relationships among genes which are difficult to infer using other methods. 

Furthermore, graph database is a growing technology these days and, in some cases,   the shortest path implementations is already at their core (for example, Neo4j \cite{Neo4j}). It is natural therefore to investigate the results in the context of this development, in order to identify possible improvements in performance gaps.




\nocite{*}
\bibliographystyle{splncs03}
\bibliography{references}

\end{document}